\documentclass[journal]{IEEEtran}
\usepackage{cite}

\ifCLASSINFOpdf
\usepackage[pdftex]{graphicx}
\else
\fi
\usepackage{amsmath}
\usepackage{color}

\usepackage{algorithm} 
\usepackage{algorithmic} 
\usepackage{subfig}
\usepackage{multirow}
\usepackage{amssymb}

\usepackage{url}

\usepackage{stfloats}

\usepackage{xcolor}
\usepackage{xpatch}

\makeatletter
\ExplSyntaxOn
\cs_new:Npn \bibColoredItems #1#2
{
	\clist_map_inline:nn {#2} { \cs_new:cpn {bib@colored@##1} {#1} } 
}
\ExplSyntaxOff

\newcommand\bib@setcolor[1]{%
	\ifcsname bib@colored@#1\endcsname
	\expandafter\color\expandafter{\csname bib@colored@#1\endcsname}
	\else
	\normalcolor
	\fi
}

\xpatchcmd\@bibitem
{\item}
{\bib@setcolor{#1}\item}
{}{\fail}

\xpatchcmd\@lbibitem
{\item}
{\bib@setcolor{#2}\item}
{}{\fail}
\makeatother

\begin{document}
	%
	\title{RELIC-GNN: Efficient State Registers Identification with Graph Neural Network for Reverse Engineering}
	%
	%
	%
	
	\author{Weitao~Pan~\IEEEmembership{Member,~IEEE, }
		Meng~Dong, 
		Zhiliang~Qiu,
		Jianlei~Yang~\IEEEmembership{Senior Member,~IEEE, } 
		Zhixiong~Di~\IEEEmembership{Member,~IEEE, }
		and Yiming~Gao
		\thanks{W. Pan, M. Dong, Z. Qiu and Y. Gao are with the State Key Laboratory of Integrated Service Networks, Xidian University, Xi'an, China (e-mail: mdong@stu.xidian.edu.cn).}
		\thanks{J. Y is with School of Computer Science and Engineering, Beihang University, Beijing, China (e-mail: jianlei@buaa.edu.cn).}
		\thanks{Z. Di is with School of Information Science and Technology, Southwest Jiaotong University, Chengdu, China (e-mail:	dizhixiong2@126.com).}}

	\maketitle

	\begin{abstract}
		Reverse engineering of gate-level netlist is critical for Hardware Trojans detection and Design Piracy counteracting.
		The primary task of gate-level reverse engineering is to separate the control and data signals from the netlist, which is mainly realized by identifying state registers with topological comparison.
		However, these methods become inefficient for large scale netlist.
		In this work, we propose RELIC-GNN, a graph neural network based state registers identification method, to address these issues.
		RELIC-GNN models the path structure of register as a graph and generates corresponding representation by considering node attributes and graph structure during training.
		The trained GNN model could be adopted to find the registers type very efficiently.
			Experimental results show that RELIC-GNN could achieve 100\% in recall, 30.49\% in precision and 88.37\% in accuracy on average across different designs, which obtains significant improvements than previous approaches.
		
	\end{abstract}
	

	%
	\IEEEpeerreviewmaketitle

	\section{Introduction}
	%
	%
	%
	%
	\IEEEPARstart{I}{n} recent years, aiming to reduce cost and time-to-market for chips, outsourcing design, manufacturing and testing services globally are becoming more and more prevalent \cite{01,21}. 
	Therefore, modern ICs pass through many hands from designers to manufacturers, which introduce many security threats such as insertion of Hardware Trojans or IP theft. 
		Hardware reverse engineering is a promising method to ensure the trustworthiness of the fabricated chips by establishing a better understanding of unknown circuits. 
	
	Reverse engineering usually consists of two stages: netlist recovery and netlist analysis \cite{02}. 
	Netlist analysis stage aims to abstract high-level information from the recovered netlist so that it is understandable by human engineers.	
		In most scenarios, it is assumed that no information except for gate-level description.
	The netlist usually consists of two parts: data-path block and control logic. 
	Data-path block usually implements some generic functions and is composed of basic components such as adder, multiplier, etc.
	On the other hand, control logic is usually designed for a specific functionality, and usually be modeled by finite state machines (FSM).
 		It is critical and most important to correctly extract FSM in a netlist for the function of control logic can be restored by it \cite{03,04,05}.
 		Reverse engineering the control logic involes the identification of the state regesters followed by a FSM extraction step.
		However, the recovered netlist is usually flattened as a mixture with data-path block and control logic, which brings great trouble for netlist analysis. 
		Therefore, correct identification of the state registers is instrumental in the extraction of a correct FSM \cite{06}.
		Many techniques have been proposed then to identify state registers in a recovered netlist \cite{06, 07, 08,09}.
	
	The accuracy and runtime of state register identification methods are significant as they determine whether FSM can be correctly identified within an acceptable time by FSM extraction algorithm. 
	However, to the best of our knowledge, most state register identification methods have long runtime. 
		Given this situation, we propose an efficient and effective state registers identification method based on graph neural network, which can fast classify registers in the arbitrary unlabeled netlist.
		By mapping the path structure of register to directed graph, the register classification can be translated into a graph classification problem.
		In RELIC-GNN, we model the netlist as a directed graph and extract the path structure for every register in the netlist.
		These path structures are then feed to the GNN model to discriminate between state and data registers.
		With the trained model, the features of the register can be obtained based on the path structure.
	Finally, the registers are grouped into state and data registers by clustering. 
	The main contributions of our work include:	
	\begin{enumerate}
		\renewcommand{\labelenumi}{\textbullet}
		\item 
			We model the netlist as a directed graph and translate the register identification into a graph classification problem by efficiently extracting path structure and features for each register.
		\item 
			We propose a reverse engineering method, RELIC-GNN, to identify state registers in a netlist by adopting state-of-the-art graph neural networks (GNN).
	\end{enumerate} 
	
	The rest of paper is organized as follows. 
	Section II presents our motivation and related works.
	Section III describes the design details of our RELIC-GNN model.
	Section IV illustrates the experimental results. 
	Section V concludes the paper.
	
	\begin{figure*}[!t]
	\centering{\includegraphics[width=\linewidth]{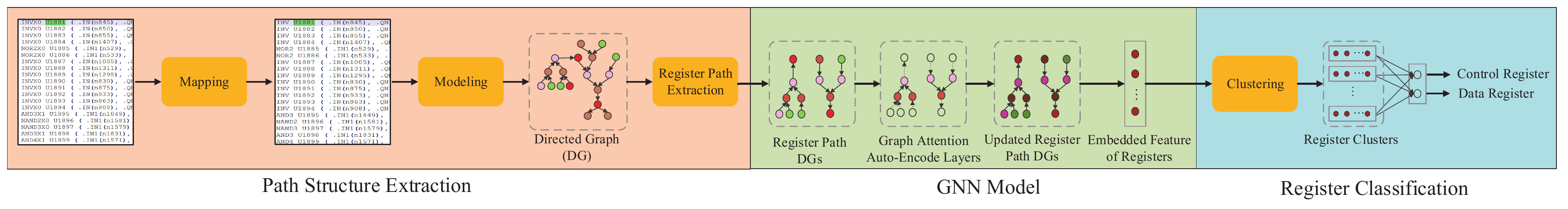}}
	\caption{An overview of proposed RELIC-GNN. Register path extraction stage first converts input netlist to technology independent netlist, and then models it as a directed graph and extracts path structure using backward breadth-first search. 
		Then, the features of registers are generated with the GNN model. 
		Finally, these registers are clustered as different groups.}
	\label{fig2}
\end{figure*}
	
	\section{Motivation and Related Work}
	
	The aim of control logic reverse engineering is to extract FSM from a gate-level netlist \cite{06}.
	FSM is the brain of design and is utilized to generate control signals and schedule the function units. 
	An FSM usually consists of a set of state registers with transitions between them. 
	On each clock cycle, the FSM is in exactly one state and moves into a new state based on the input combinations and the current state.
		Therefore, a register can be identified as a state register if it has a feedback path \cite{03}.
		However, it is not always accurate due many data-path blocks contain registers including feedback paths.
		To improve the identification accuracy, Shi et al. \cite{04} eliminates all identified candidate state registers which do not affect a control signal, such as the select line for a multiplexer or the enable signal for a register.
	
	
	RELIC \cite{06} identifies control signals based on the similarity of the fan-in structure. 
		It first pre-processes the netlist such that it contains only AND, OR, and INV gates, to reduce the structure complexity.
	Then, RELIC computes a Pair Similarity Score (PSS) for every two registers according to the fan-in structural similarity and the type of input gate. 
	Finally, the registers are classified according to the PSS. 
	However, its efficiency and accuracy are not good enough.
	fastRELIC \cite{07} improves RELIC from two aspects. 
	Firstly, fastRELIC computes PSS of the signal and its inverse, and selects the larger value as the final PSS. 
	Secondly, fastRELIC introduces the idea of clustering in the PSS calculation stage, which greatly reduces the number of computations of PSS, thereby improving its efficiency.
	REBUS \cite{08} is another improvement method of RELIC, which aims at reconstructing the data paths of a netlist. 
	It first randomly selects a node, and then calculates PSS with ungrouped nodes. 
	The nodes with PSS higher than the threshold are combined into a new word.	
		ReIGNN\cite{09} combines GraphSAGE with structural analysis to classify the registers.
		It translates register classification task into a node classification problem by mapping the circuit to graph.
		RelGNN first train the GraphSAGE that processes graph to discriminate between state and data registers in netlist.
		Finally, the misclassifications of GraphSAGE are rectified by structural analysis of the netlist.
	
	
	The path structures can be modeled as directed graphs, which are non-Euclidean spatial data structures. 
	Therefore, the task of identifying state registers can be redefined as a graph classification problem, and machine learning is a powerful tool for such problem. 
	As it is difficult and even unrealistic to obtain exact category of registers from the netlist, unsupervised learning model is more suited to this problem.
	Auto-encoders are popular in unsupervised learning due to their ability to capture complex relationships between attributes of the input through non-linear layers \cite{10,11,12}. 
	In addition, GNN is a powerful neural network architecture mainly used for machine learning of non-Euclidean spatial data structures. 
	To capture node relationships in the non-Euclidean spatial data structures, the graph auto-encoder with GNN has been proposed \cite{13}. 
	The combination of auto-encoders and GNN has opened new opportunities for netlist analysis, which is the main motivation of this paper.

	\section{Methodology}
		The work flow of RELIC-GNN is shown in Fig. 1, consisting of path structure extraction, GNN model and register classification.
		RELIC-GNN extracts the path structure of register and models it using GNN, and then classifies registers by embedding features.

	\subsection{Path Structure Extraction}
	
		In IC design, cells with the same logical function usually have different descriptions because they vary in timing, area, and power, which makes it more difficult to analyze the netlist function.
	Therefore, in order to reduce the complexity of the algorithm, we first learn the existing technology libraries and build the technology-independent cell model. 
	Then, the cells in netlist are replaced by cells in technology-independent cell model with the same function.  
	
	After mapping cell to technology-independent cell model, we model the technology independent netlist as a directed graph $G=(V,E,F)$, where $V$ is the set of cells, $E$ is the set of data dependencies between the cells, and $F$ is the set of unique identifiers of cells.
	$V= \{v_1,v_2,...,v_n\}$, where $v_i$ denotes cell name such as xor2, and2, etc.
	$E = \{e_{ij}\}$, where $e_{ij}$ is 1 if $v_i$ is the fan-in cell of $v_j$ and 0 otherwise. 
	$F= \{f_1,f_2,...,f_n\}$, where $f_i$ is the feature vector of $v_i$. 
		The feature vector $f$ of a node contains the \textit{cell type}, \textit{in-degree} and \textit{out-degree}  information. 
		The \textit{cell type} is represented by a one-hot encoding vector which dimension is associated with the technology-independent cell model built before.
		And the \textit{in-degree} and \textit{out-degree} is the number of incoming and outgoing neighbors.
	A processed gate-level netlist and its directed graph are exemplified in Fig. 2.
	\begin{figure}[!h]
		\centering{\includegraphics[width=\linewidth]{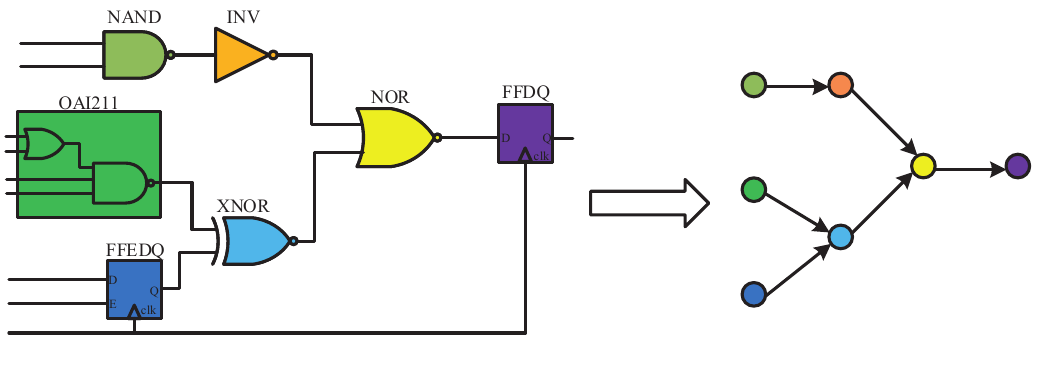}}
		\caption{Translating technology independent netlist as a graph. The node color in the graph represents the cell type.}
		\label{fig3}
	\end{figure}
	
	Algorithm 1 shows the algorithm for path structure extraction based on the directed graph. 
	The algorithm attempts to find the path structure of register by executing backward breadth-first search. 
		Since the control logic in a design is in charge of sending critical signals and controlling the datapath, the path of it usually is short to avoid high delay.
		Consequently, a path structure with a depth of 6 can distinguish control registers and data registers in most cases. 
	In addition, to improve the run-time efficiency, the algorithm terminates immediately if another register appears in the search, as shown in lines 9-11 of the algorithm 1.
	
	\begin{algorithm} [h]\small 
		\caption{Path Structure Extraction} 
		\label{alg1} 
		\begin{algorithmic}[1] 
			\REQUIRE ~~\\
			\textbf{start\_node}: The start gate of sub-circuits extraction. \\ 
			\textbf{predecessor\_set}: The predecessor set of each gate in netlist.\\ 
			\textbf{walk\_length}: The depth of search.
			
			\ENSURE ~~\\ \textbf{walk\_nodes}: The gates in sub-circuits.
			
			\STATE walk\_nodes $\leftarrow$ [[start\_node]]
			\STATE walk $\leftarrow$ []
			\FOR {$i$ $\textless$ walk\_length}
			\STATE Is\_queit $\leftarrow$ False
			\FOR {node $\in$ walk\_node[-1]}
			\STATE predecessor\_node $\leftarrow$ predecessor\_set
			\FOR {pre\_node $\in$ predecessor\_node}
			\STATE walk.add(pre\_node)
			\IF {pre\_node is start\_node \textbf{or} pre\_node is register}
			\STATE Is\_queit $\leftarrow$ True
			\ENDIF
			\ENDFOR
			\ENDFOR
			\IF {Is\_queit is True}
			\STATE break
			\ELSE 
			\STATE $i = i + 1$
			\STATE walk\_nodes.add(walk)
			\ENDIF
			\ENDFOR
		\end{algorithmic} 
	\end{algorithm}

	\subsection{GNN Model}
	
	In RELIC-GNN, we leverage GNN model to learn the complex, non-linear relationship between the path structure and attributes of registers.
	Our architecture is inspired by the Graph Attention Auto-Encoders Network (\texttt{GATE}) \cite{13}, which combines the GNN with auto-encoders.
	As shown in Fig. 3, the \texttt{GATE} network consists of two phases: encoding and decoding. 
	In the encoding phase, the "gate embedding" process and stacked Graph Attention Network (GAT) are used to encode the graph. 
	Each encoding layer generates a new node representation based on the relevance of the node, using the domain representation of the node. 
	It involves two functions: computing relevance coefficients, given by Eq.\eqref{eq1}, and generating presentation of nodes, given by Eq.\eqref{eq2} and \eqref{eq3}:

		{\footnotesize \begin{equation}
			\label{eq1}
			e_{ij}^{(k)}\!=\!\operatorname{Sigmoid}\left(\!\mathbf{v}_{s}^{(k)^{T}}\!\sigma\left(\!\mathbf{W}^{(k)}\!\mathbf{h}_{i}^{(k-1)}\right)\!+\!\mathbf{v}_{r}^{(k)^{T}}\!\sigma\left(\!\mathbf{W}^{(k)}\!\mathbf{h}_{j}^{(k-1)}\!\right)\!\right),
		\end{equation}}	
		{\footnotesize \begin{equation}
			\label{eq2}
			\alpha_{i j}^{(k)}=\frac{\exp \left(e_{i j}^{(k)}\right)}{\sum_{l \in N_{i}} \exp \left(e_{i l}^{(k)}\right)},
		\end{equation}}
		{\footnotesize \begin{equation}
			\label{eq3}
			\mathbf{h}_{i}^{(k)}=\sum_{j \in N_{i}} \alpha_{ij}^{(k)} \sigma\left(\mathbf{W}^{(k)} \mathbf{h}_{j}^{(k-1)}\right),
		\end{equation}}
	where $\mathbf{W}^{(k)} \in \mathbb{R}^{d^{(k)} \times d^{(k-1)}}$, $\mathbf{v}_{s}^{(k)} \in \mathbb{R}^{d^{(k)}}$, and $\mathbf{v}_{r}^{(k)} \in \mathbb{R}^{d^{(k)}}$ are trainable parameter of the $k^{th}$ encoding layer, $\sigma$ denotes the activation function, $\alpha_{i j}^{(k)}$ is the normalization of $ e_{ij}^{(k)}$ and represents the relative relevance of neighboring node $j$ to node $i$ in the $k^{th}$ encoder layer. Essentially, the representation of node $i$ in layer $k$ is generated by combining the previous neighboring nodes features $\mathbf{h}_{i}^{(k-1)}$ with $\alpha_{i j}^{(k)}$. The output of the last encoding layer is the final node representations.
	
	In the decoding phase, the decoder contains the same number of layers as the encoder. 
	And each decoding layer reconstructs the representation of node based on the final node representations.
	
	\begin{figure}[!h]
		\centering{\includegraphics[width=\linewidth]{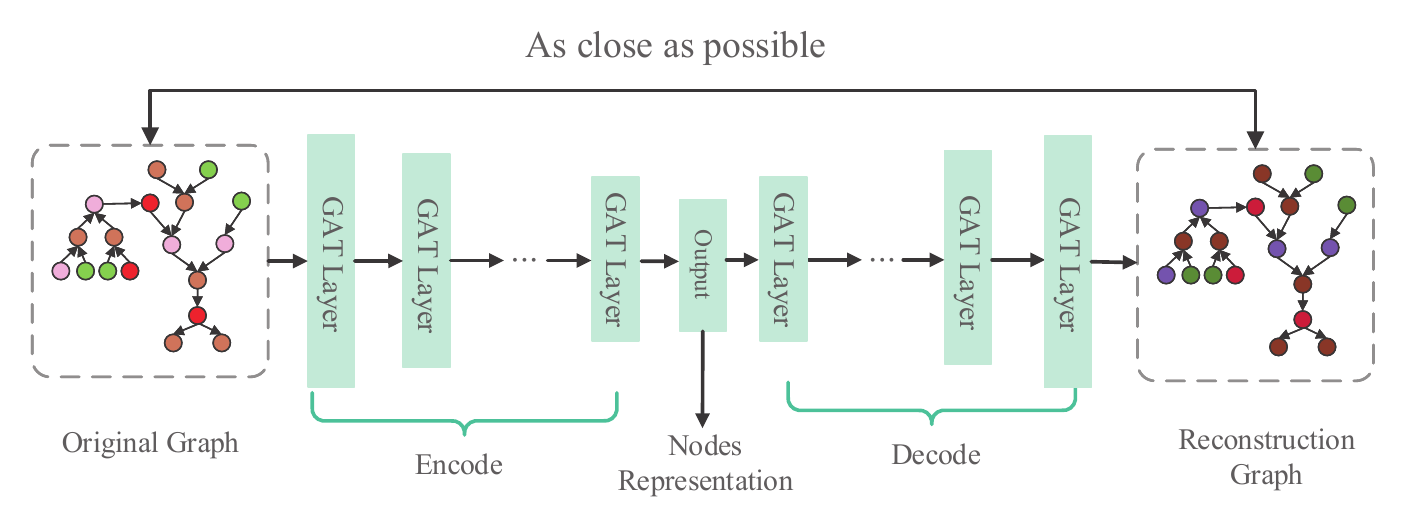}}
		\caption{The framework of \texttt{GATE}.}
		\label{fig4}
	\end{figure}
	
\begin{table}[b]
	\label{table1}
	\caption{
			Benchmarks overview, sorted by total number of registers.
	}
	\centering
	\begin{tabular}{l|rrr}
		\hline\hline
		Designs & State Registers & Registers & Gates \\
		\hline
		FSM$\ddagger$                 &4      &10        &84       \\
		b10$\star$                &4      &17        &107      \\
		b08$\star$                 &2      &21        &89       \\
		b09$\star$                 &2      &28        &108      \\
		gpio$\ddagger$                &4      &44        &216      \\
		b04$\star$                 &2      &66        &307      \\
		MEM$\ddagger$                 &4      &72        &822      \\
		b14$\star$                 &1      &215       &2245     \\
		spi\_axi$\ddagger$            &4      &547       &1314     \\
		uart$\dagger$                &7      &604       &2270     \\
		siphash$\diamondsuit$             &8      &789       &4035     \\
		mem\_controller$\dagger$     &66     &1051      &4158     \\
		alto32$\dagger$              &6      &1246      &6288     \\
		sha1$\diamondsuit$                &2      &1525      &7328     \\
		gcm\_aes$\dagger$            &10     &1696      &19231    \\
		lightweight\_8051$\dagger$   &4      &2642      &9332     \\            
		aes$\diamondsuit$                 &14     &2987      &16011    \\ 
		cr\_div$\dagger$             &3      &4171      &15326    \\
		escda$\S$               &10     &113432    &1318598  \\
		\hline\hline
	\end{tabular}
\end{table}
	To obtain more accurate representation of the nodes, multi-headed attention mechanism is used in aggregating the neighboring features of nodes, and the number of encoder layers is set as 3.

	\subsection{Register Classification}	
	We leverage clustering technology to reduce the complexity and improve the efficiency of registers classification.
	First, all register nodes are marked as candidate register nodes (CRNs) and placed into a candidate group (CG). 
	The follow process is iterated until there is only one or no CRN in CG. 
	In the $i^{th}$ iteration, a starting register node (SRN) is randomly fetched from CG and put into an empty set $S_i$. 
		Then we calculate the absolute value of the node feature difference (FDV) between SRN and all other nodes in CG.
		The nodes in CG whose FDV is less than the threshold $T1$ are fetched and put into $S_i$, while others remain.
	In order to separate the control register and data register as much as possible, $T1$ is set to $1 \times 10^{-3}$.
	
	After grouping, we count the number of nodes in $S_i$. 
	When the number is less than a threshold $T2$, all nodes in the $S_i$ are classified as state registers, otherwise as data registers. 
	Since the minimum multi-bit data path in the chip design is 4 bits, $T2$ is set to 4.
	
\begin{table}[t]
	\caption{
			GNN Configuration.
	}
	\label{table2}
	\centering
	\begin{tabular}{ll|ll}
		\hline\hline
		\multicolumn{2}{c}{\textbf{Architecture}} & \multicolumn{2}{c}{\textbf{Training Setings}} \\
		\hline
		Encoder Layer \#1 & $\left[|f_i|,64\right]$  & Optimizer &Adam\\
		Encoder Layer \#2 & $\left[64,64\right]$    &Learning Rate &0.01 \\
		Encoder Layer \#3 & $\left[64,1\right]$      & Weight Deacy&5e-4 \\
		Decoder Layer \#1 & $\left[1,64\right]$     & NO. of Epochs&200 \\
		Decoder Layer \#2 & $\left[64,64\right]$    & Gradient Clipping&5 \\
		Decoder Layer \#3 & $\left[64,|f_i|\right]$& & \\
		\hline\hline
	\end{tabular}
\end{table}

\begin{table*}[hb]
	\caption{
			Evaluation results comparison among fastRELIC \cite{07} ReIGNN \cite{09}, and RELIC-GNN (this work) on $Recall$, $Precision$, and $Accuracy$.
	}
	\label{table3}
	\centering
	\begin{tabular}{l|rrr|rrr|rrr}
		\hline\hline
		\multirow{2}{*}{Design} & \multicolumn{3}{c}{$Recall$} & \multicolumn{3}{c}{$Precision$}&\multicolumn{3}{c}{$Accuracy$} \\
		\cline{2-10}
		&fastRELIC &ReIGNN & RELIC-GNN &fastRELIC &ReIGNN & RELIC-GNN &fastRELIC &ReIGNN & RELIC-GNN \\ 
		\hline
		FSM$\ddagger$               &100\%     &100\%    &100\%   &80\%     &100\%      &100\%    &64.29\%   &71.43\%     &71.43\%   \\
		b10$\star$                  &75\%      &100\%    &100\%   &25\%     &44.44\%    &50\%     &36.84\%   &60.87\%     &61.90\%   \\
		b08$\star$                  &100\%     &100\%    &100\%   &13.33\%  &25\%       &22.22\%  &34.78\%   &65.22\%     &60.87\%   \\
		b09$\star$                  &100\%     &100\%    &100\%   &40\%     &28.57\%    &40\%     &83.33\%   &76.67\%     &83.33\%   \\
		gpio$\ddagger$              &100\%     &75\%     &100\%   &26.67\%  &33.33\%    &66.67\%  &68.75\%   &80.43\%     &87.50\%   \\
		b04$\star$                  &100\%     &100\%    &100\%   &11.76\%  &18.18\%    &22.22\%  &75\%      &83.82\%     &86.76\%   \\
		MEM$\ddagger$               &100\%     &100\%    &100\%   &30.77\%  &30.77\%    &44.44\%  &82.89\%   &82.89\%     &88.16\%   \\         
		b14$\star$                  &100\%     &100\%    &100\%   &8.33\%   &4.35\%     &3.85\%   &94.44\%   &89.35\%     &87.96\%   \\         
		spi\_axi$\ddagger$          &100\%     &100\%    &100\%   &22.22\%  &7.27\%     &33.33\%  &96.75\%   &90.07\%     &97.83\%   \\
		uart$\dagger$               &100\%     &100\%    &100\%   &21.21\%  &2.85\%     &53.81\%  &94.6\%    &89.74\%     &97.87\%   \\
		siphash$\diamondsuit$       &100\%     &100\%    &100\%   &6.12\%   &3.37\%     &15.79\%  &93.81\%   &88.76\%     &97.60\%   \\
		mem\_control$\dagger$       &74.24\%   &96.97\%  &100\%   &27.07\%  &29.09\%    &25.29\%  &84.33\%   &81.43\%     &78.10\%   \\
		alto32$\dagger$             &100\%     &66.67\%  &100\%   &4.23\%   &2.99\%     &4.92\%   &88.66\%   &89.26\%     &90.26\%   \\
		sha1$\diamondsuit$          &100\%     &100\%    &100\%   &2.70\%   &1.20\%     &13.33\%  &95.15\%   &89.06\%     &99.02\%   \\
		gcm\_aes$\dagger$           &100\%     &100\%    &100\%   &4.17\%   &5.75\%     &29.41\%  &85.99\%   &89.72\%     &98.02\%   \\
		lightweight\_8051$\dagger$  &100\%     &75\%     &100\%   &1.74\%   &1.14\%     &3.33\%   &91.31\%   &90.05\%     &95.46\%   \\
		aes$\diamondsuit$           &100\%     &100\%    &100\%   &6.06\%   &2.81\%     &22.86\%  &95.59\%   &90.48\%     &98.83\%   \\
		cr\_div$\dagger$            &100\%     &100\%    &100\%   &3.71\%   &0.77\%     &27.27\%  &98.06\%   &90.61\%     &99.74\%   \\
		escda$\S$                   &NA        &NA       &100\%   &NA       &NA         &0.58\%   &NA        &NA          &98.47\%   \\
		\hline
		\textbf{average}            &97.18\%   &95.20\%  &100\%   &18.62\%  &18.99\%    &30.49\%  &81.37\%   &83.30\%     &88.37\%  \\
		\hline
		\hline
	\end{tabular}
\end{table*}

	\subsection{Complexity Analysis}
	
	As shown in Fig. 1, the complexity of RELIC-GNN consists of three parts. 
	Assuming that the number of registers in netlist is $R$, we analyze the upper bound of complexity of RELIC-GNN algorithm in terms of register path structure extraction, GNN model and register classification.
	
	
	\subsubsection{Complexity Analysis of Path Structure Extraction}	
	During the netlist mapping step, the fan-in information of each node is stored in the form of dictionary. 
	Therefore the time complexity of finding the fan-in nodes of a cell is only $\mathcal{O}(\text{1})$. 
	In addition, the fan-in nodes of a cell are usually not more than 4 in IC design.
	The path structure extraction must be repeated for all registers, yielding the following factor (MF1):	
	
{\footnotesize 	\begin{equation}
		\label{eq5}
		\text{MF1}=\left(\sum_{i=1}^{\text{walk\_length-1}} 4^{i}\right) \times R
	\end{equation}}	
	\subsubsection{Complexity Analysis of GNN Model}
	For GNN model, the time complexity is $\mathcal{O}\left(N F F^{\prime}+E F^{\prime}\right)$, where $N$ is the number of nodes in the graph, $F$ is the dimension of the initial features of each node, $F'$ is the dimension of the output features of each node, and $E$ is the number of edges in the register path structure.  
	The GNN model must be applied to all extracted path structures yielding the following factor (MF2):		
		
		{\footnotesize \begin{equation}
			\label{eq6}
			\text{MF2}=\left(\sum_{i=1}^{\text{walk\_length-1}} 4^{i} \times F F^{\prime}+\sum_{i=1}^{\text{walk\_length-1}} 4^{i} \times F^{\prime}\right) \times R
		\end{equation}}
	
	\subsubsection{Complexity Analysis of Register Classification}
	In register classification, it is necessary to calculate the FDV between registers. 
	In the worst case, FDV calculation is needed for every two registers. 
	Therefore, the upper bound of time complexity is shown in Eq. \eqref{eq7}.
	
		{\footnotesize \begin{equation}
			\label{eq7}
			\mathcal{O}\left(\frac{R(R-1)}{2}\right)=\mathcal{O}\left(R^{2}\right)
		\end{equation}}
	
	Consequently, the upper bound complexity for the RELIC-GNN proposed in this paper is shown in Eq. \eqref{eq8}.	
		{\footnotesize \begin{equation}
			\label{eq8}
				\mathcal{O}(\text { RELIC-GNN}) =\mathcal{O}(\text{MF1})\!+\!\mathcal{O}(\text{MF2})\!+\!\mathcal{O}\left({R}^{2}\right) \\
				=\mathcal{O}\left({R}^{2}\right)
		\end{equation}}

	\section{Experimental Results}
		In this section, we evaluate the performance of RELIC-GNN and compare with two baseline: fastRELIC and ReIGNN.
	We conduct GNN model training and inference experiments on a computer with 	an NVIDIA 2080Ti GPU, an Intel(R) Core(TM) i5-7500, and 16-GB memory.
	RELIC-GNN and our baseline fastRELIC and ReIGNN both run on the same netlist,  synthesized using Synopsys Design Compiler \cite{14}, together with the SMIC 28nm cell library.
	
	
		Benchmarks used in this paper are listed in Table I, containing the ITC99 benchmark ($\star$) \cite{15}, designs in Opencore ($\dagger$) \cite{16}, the secworks github repository ($\diamondsuit$) \cite{17}, the 32-bit RISC-V processor in github ($\ddagger$) \cite{18}, and the elliptic curve digital signature algorithm in github ($\S$) \cite{19}. 
	The number of registers and gates of netlist are derived from synthesized netlist, whereas the number of state registers is deduced by RTL description.

		The GNN model consists of 3 encoder layers and 3 decoder layers.
		The first encoder layer maps the high dimension input node features to a lower dimension space (we chose 64 dimensions). 
		And the gradient descent optimizer Adam \cite{20} is used to update the parameters in the GNN model by introducing an unsupervised loss function.
		Table II defines the GNN model as well as the training architecture.

	
		To evaluate the performance, we perform a leave-one-out cross-validation. 
		GNN models are trained on 18 of the 19 total netlist, leaving 1 netlist outside of training data set to be the test set data. 
		In this way, we may verify the transferability of the proposed GNN architecture on all 19 circuits.
		We use true positives (TP), true negatives (TN), false positives (FP), and false negatives (FN) to calculate the evaluation metrics. 
		The evaluation metrics are recall, precision and accuracy.
		The calculation formulas are expressed as follows: $Recall=TP/(TP+FN)$, $Precision=TP/(TP+FP)$, and $Accuracy=(TP+TN)/(TP+TN+FP+FN)$.
	
	In our experiments, we did not use the reported numbers of Brunner et al. \cite{07} and Chowdhury S D et al. \cite{09}, as the operating environment affects the efficiency of algorithm. 
	In order to have a fair comparison, we tried to obtain the executables of their implementations. 
	Unfortunately, we were not able to get the implementation of them. 
	We used our own implementation of their approach in experiments.	
		In fastRELIC \cite{07}, the user need to set the three threshold values $T1$, $T2$, and $T3$.
		In our experiments, we consider parameter configuration is $T1=0.7$, $T2=0.5$, and $T3=4$.
	
	
		Table III shows the comparison results with fastRELIC \cite{05} and ReIGNN \cite{09}.
		It can be seen that RELIC-GNN outperforms the fastRELIC and ReIGNN in recall, precision and accuracy.
		It should be noted that enhancing both recall and precision is important for state registers classification become we aim to minimize both FN and FP. 
		Therefore, the proposed RELIC-GNN demonstrates a more balanced classification performance.
		In addition, for the escda netlist, proposed RELIC-GNN identify the registers within 16 hours, and fastRELIC and ReIGNN cannot get classification result in 24 hours.

	\section{Conclusion}
	
	This paper applies GNN to the field of reverse engineering analysis, and proposes a reverse engineering method RELIC-GNN. 
		Experiment results show that proposed RELIC-GNN achieves higher recall and accuracy, especially in precision, with respect to previous classification methods.
		Theoretical analysis and experiments show that RELIC-GNN can support the state register identification of millions of gate-level netlist.
		Future work includes further improve the precision and reduce complexity of proposed architecture.
	
	\hspace*{\fill} 

	\ifCLASSOPTIONcaptionsoff
	\newpage
	\fi

	
	
	%
	%
	%
	
	\bibliographystyle{IEEEtran}
	
	
	
	\bibliography{IEEEabrv,reference.bib}

	%
	
	
	
	
	
	
	

\end{document}